\documentclass[final,3p,times,twocolumn,authoryear]{elsarticle}

\usepackage{amssymb}
\usepackage{color}
\usepackage{graphicx}
\usepackage{multirow}
\usepackage{caption}
\usepackage{parskip}
\usepackage{subfigure}
\usepackage[encapsulated]{CJK} 
\usepackage{float} 
\usepackage{listings}
\usepackage{xcolor}
\usepackage[authoryear]{natbib}
\definecolor{Blue}{rgb}{0,0,1}
\definecolor{Green}{rgb}{0,0.5,0}
\definecolor{Red}{rgb}{0.64,0.08,0.08}
\journal{arxiv} 

\begin{document}

\begin{frontmatter}



\title{Judgment2vec: Apply Graph Analytics to Searching and Recommendation of Similar Judgments}


%
\author{Hsuan-Lei Shao}

\affiliation{organization={Graduate Institute of Health and Biotechnology Law, Taipei Medical University}
            }

\begin{abstract}
In court practice, legal professionals rely on their training to provide opinions that resolve cases, one of the most crucial aspects being the ability to identify similar judgments from previous courts efficiently. However, finding a similar case is challenging and often depends on experience, legal domain knowledge, and extensive labor hours, making veteran lawyers or judges indispensable. This research aims to automate the analysis of judgment text similarity. We utilized a judgment dataset labeled as the "golden standard" by experts, which includes human-verified features that can be converted into an "expert similarity score." We then constructed a knowledge graph based on "case-article" relationships, ranking each case using natural language processing to derive a "Node2vec similarity score." By evaluating these two similarity scores, we identified their discrepancies and relationships. The results can significantly reduce the labor hours required for legal searches and recommendations, with potential applications extending to various fields of information retrieval.

\end{abstract}




\end{frontmatter}

\section{Research Background} 

It is crucial to know similarity judgment which means locating legal texts and related applications, including recommendation of legal judgments. In the court process, judges or lawyers need to find former similar cases as a clue to help them to enforce their discourse or to decide better. However, finding “a similar” case is a difficult task which is often based on their experience, academically trained knowledge, and labor hours. No royal road in this field. 

Strictly speaking, There are no total indifference pairs of judgments in the real world -- like “people can not step twice into the same river.” However, we still have to use the former case instead of by our ourselves' thinking to persuade other members in the court--can this skill be quantified or calculated? Therefore, this paper try to use digital methods to analyze the similarity between judgment texts. Firstly, we used the judgment dataset, which experts label as the "golden standard". The labeled dataset has human-eye-checked features. Therefore we could calculate the cosine similarity between sample pairs in this dataset. Then we can take this similarity matrix as a primary standard. In this stage, we take 124 cases of elder alimony lawsuits in Taiwan.

In detail, why “similar judgments” are essential to legal practitioners? In theory, the legal practice of the “continental law system” is: step 1. Finding possible proper legal article; step 2. Observe whether the facts can fit the legal constituent elements, and step 3. Then follow the legal constituent elements to produce legal effects. The two parties’ legal disputes often occur in steps 1 and 2. Therefore, in practice, they will support their legal views according to “similar judgments,” At the same time, they can expect their legal result. Therefore, even though each case is “independently judged” in legal philosophy, it is still necessary to refer to the previous examples in practice. Even if there is a database of legal judgments, only one or several keywords can be used, and each user must read each search results with human eyes to select the similar judgment he wants—not like a shopping website. They can provide recommended products or similar products. Therefore, a senior lawyer can be very valuable because he/she has long-term relevant experience and can accurately input keywords or quickly judge an important legal issue or similar judgments.

Another reason why our research is necessary is that most studies on similar judgments use judgment corpora from "common law systems" those often directly cite previous judgments (jurisprudence). Therefore, obtaining the relationship between judgments is relatively simple than using the continental law judgment corpus for research. Because of the different legal concepts, there are few references to each other—so we must use natural language processing skill to extract features as a similarity. Judgmental links can provide more information as a basis for similar classification. To make a relationship without original citation is our technological innovation.

The second is that we compared the “similarity” considered by the algorithm with the “similarity” identified by legal experts, which is relatively rare in other studies. Therefore, our contribution is to use this algorithm to expand to other types of legal judgments, which can automatically bypass the labeling stage and obtain similar judgment results on a larger scale, which can provide a new application possibility for legal search.

\section{Related Researches } 
A related paper to challenge similar judgments from the information science field is “Similarity analysis of legal judgments” in 2011\cite{r1}. This paper tried to find similar legal judgments by extending the popular techniques used in information retrieval and search engines. Legal judgments are complex and refer to other judgments. They have analyzed all-term, legal-term, co-citation, and bibliographic coupling-based similarity methods to find similar judgments. The experimental results show that the legal-term cosine similarity method performs better than the all-term cosine similarity.\cite{r1} This paper hints the "cosine similarity" can be a index for similar judgments, and we take this idea. However, compared with experts’ review, the machine learning method shows that the all-term cosine and co-citation similarity methods are inefficient in finding similar judgments. The legal terms cosine and bibliographic coupling similarity methods efficiently find similar legal judgments.\cite{r1} However, this paper is by the “common law” system. The judgment content itself has several “citations” from the former judgments. Therefore, it can be easy to build a relationship with each other. This method cannot fully fit into the “continental law” system the Taiwanese use. So, it is an issue that shall be solved, but we cannot follow this method.

A paper, LawNet-Viz: A Web-based System to Visually Explore Networks of Law Article References\cite{r5}, has given us the inspiration to do graph analysis, which allows the user to refine a network according to the semantic similarity of linked articles, which is captured based on sparse vector representations, topic models, word embeddings or deeply contextualized embeddings\cite{r5}. It provides us with more knowledge possibilities in this research-- almost like our ideal completion type. However, because our corpus does not have “connections between articles to define a citation or reference network,” [5]we cannot fully follow the stream of the graph representation model of the connections of the LawNet-Viz. We need an annotation research design for our continental law judgments corpus.

 Another paper, “Distinguish Confusing Law Articles for Legal Judgment Prediction” \cite{r2}. It noticed a problem: “law cases applicable to similar law articles are easily misjudged. It relies heavily on domain experts, which hinders its application in different law systems.” For the issue, This article used a graph neural network to learn subtle differences between confusing law articles automatically. Moreover, they design an attention mechanism that fully exploits the learned differences to extract compelling discriminative features from fact descriptions. \cite{r2} This article is to find the correct article for the “given fact.” Moreover, it gave us a hint: if it is possible to input a text of facts, then extract features and find the right “law article.” We may take one more step to give those features a rank to find similar judgments. Because there is still more detailed information in real-world cases, we want to make a judgment, and a recommended system for legal workers, helping them save laborers. 

A paper that has the same interest is “Methods for Computing Legal Document Similarity: A Comparative Study” (2004)\cite{r4}. It also uses Indian (common law) system judgments, analyzing the precedent citation network and measuring similarity based on textual content similarity measures. The Main point of view is to list several definitions of “similarity” and prove that similarities can be different from each other. as table1.

\begin{table}[H]
\caption{Different Method of Similarity in the same pair text[4]}
\resizebox{\linewidth}{10mm}{
\begin{tabular}{|l|l|l|l|l|l|}
\hline
Document Pair        & Expert Score & Node2Vec & FullText Similarity & \begin{tabular}[c]{@{}l@{}}Thematic \\ Similarity(Avg)\end{tabular} & \begin{tabular}[c]{@{}l@{}}Thematic \\ Similarity(Max)\end{tabular} \\ \hline
1992\_47\&1992\_76   & 0            & 0.195    & 0.188               & \textbf{0.154}                                                      & 0.571                                                               \\ \hline
1979\_110\&1989\_233 & 3            & 0.613    & 0.465               & 0.104                                                               & \textbf{0.415}                                                      \\ \hline
1953\_24\&1957\_52   & 7            & 0.234    & 0.264               & 0.377                                                               & \textbf{0.757}                                                      \\ \hline
1983\_37\&1979\_33   & 10           & 0.574    & \textbf{0.711}      & 0.209                                                               & 0.692                                                               \\ \hline
\end{tabular}
}
\end{table}
\vspace{-0.5cm} 

This paper used the expert-labeled data and Node2Vec can achieve the best proved cooperation in the prior-case
citation network based measures\cite{r4}. We use this experience in the "continental law system."

\section{Research Design} 
\subsection{Research Data and Preprocessing} 

We used 494 judgments on old-age alimony from 2013 to 2018. Most of the judgments included multiple defendants. Since most of the characteristics of old-age maintenance are to meet the minimum (or average) standard of living when multiple defendants often The relationship between a defendant and the plaintiff will affect the “sharing amount of each defendant.” We are worried that this “multiple defendants” will affect the calculation, so we only take 124 samples as the sorting set for this experiment.

From these 124 cases, we obtained as much as possible the legal elements considered by legal researchers to classify cases: \emph{“average monthly living expenses per person, monthly minimum living expenses, whether the claimant is capable of working, whether the claimant Physical and mental disabilities, whether the claimant has a disease, whether the claimant is unable to claim subsidies because of his daughter, the amount of the claimant’s annuity, the amount of other monthly income of the claimant, child support, the average taxable income of the claimant, whether the claimant has real estate, the claimant Total assets, the amount of medical expenses of the claimant, the amount requested by the claimant, the age of the claimant, the status of the claimant, the gender of the claimant, the monthly income of the relative, the average taxable income of the relative, the fixed expenditure of the relative, whether the relative has real estate, whether the relative Total assets, number of relatives, number of relatives’ brothers and sisters (existing), assets of relatives’ other siblings, age of relatives, compliance with 1118-1 items, whether the petitioner is divorced, whether the petitioner has parental rights, outstanding support The number of years of obligation, whether the claimant has abused the respondent, whether the relative has no written statement or has not appeared in court, whether the claimant has a lawyer, and whether the relative has a lawyer”, and are manually marked as our core (golden-standard) dataset. We normalize the float factor like “ the monthly income of the relative”}, because there can be a large gap in different cases. The normalize method is:
\begin{lstlisting}[language=python]
factor.y-factor.min()/factor.max()-factor.min().
\end{lstlisting}
The core data set’s function may use similar algorithms as in the paper \cite{r1}. Under the agreement of certain legally specific factors, these judgments may be similar. Secondly, we will also obtain a set of similar results by using the graph calculation, which is used to compare with the similar sets manually marked by legal experts, which is relatively close to the practice of paper \cite{r2}. Finally, we use the dataset to build a basic similarity matrix, as fig.1 below.

\begin{figure}[H] 
  \centering 
  \includegraphics[width=0.8\linewidth]{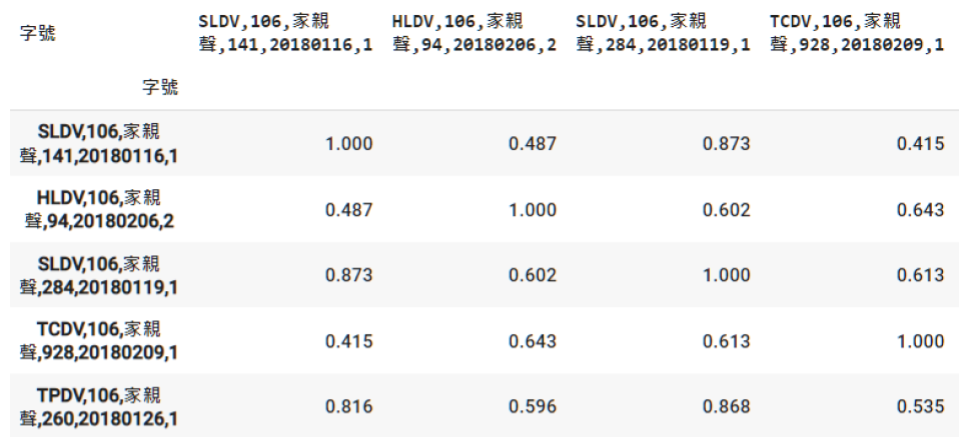} 
  \caption{Cosine Similarity Matrix by Human Label} 
\end{figure}
\vspace{-1.0cm} 

\subsection{Graph Similarity Research Flow} 

As mentioned, it is expensive to hire a expert to label dataset. Our target is to do it automatically. Then we design a primary text analysis (measuring similarity based on textual content similarity measures) to judge. We can not construct a network by citation judgment, as mentioned. We use basic information, such as law articles and district court locations. Nevertheless, it could be more information to analyze. 

 Therefore, we retrieve the content feature of the judgment and extract the judgment text words by the LDA method (Latent Dirichlet Allocation). We have tried two different LDA preprocessing: by normal DTM(document term matrix) with segmentation, and by the TF-IDF+DTM with segmentation.


\begin{figure}[H] 
  \centering 
  \includegraphics[width=1\linewidth]{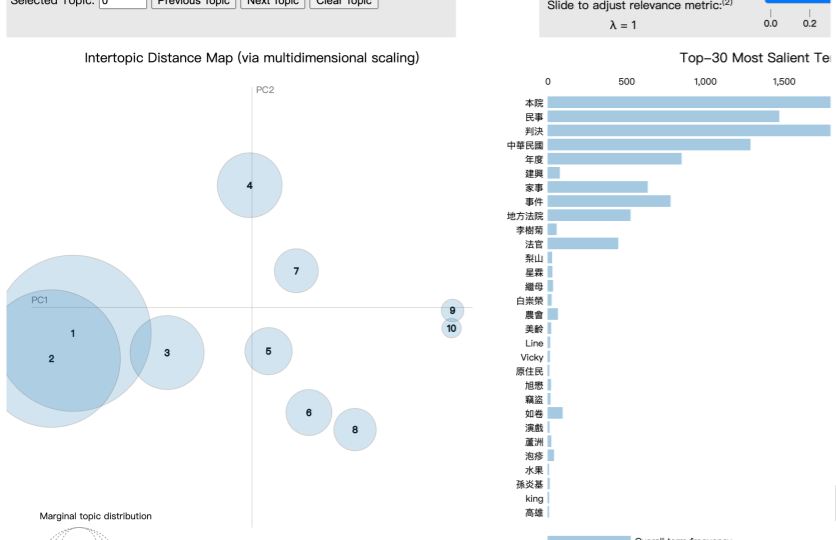} 
  \caption{Result of LDA by TF-IDF+DTM} 
\end{figure}
\vspace{-0.5cm} 

As the figures, the result of the LDA model by TF-IDF+DTM can retrieve more legal-related words. We take those “key words” as a new node to enhance new features, like those judgments and build more linkage. For example, we can add more attributes for one case as figure 3 below.

\begin{figure}[H] 
  \centering 
  \includegraphics[width=0.7\linewidth]{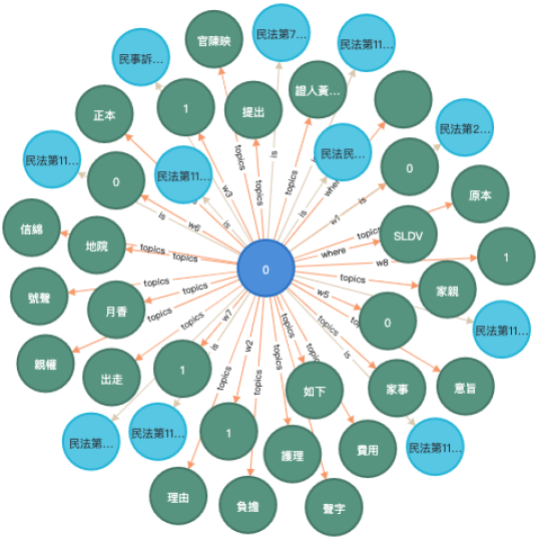} 
  \caption{Attributes Added Case Node} 
\end{figure}
\vspace{-0.5cm} 

After that preprocessing, we can cross the barrier of non-citation text and generate a relationship graph of judgments similarity (see next page: figure 4,\emph{ Judgement Relationship Network Graph with topics})

\begin{figure}[H] 
  \centering 
  \includegraphics[width=0.7\linewidth]{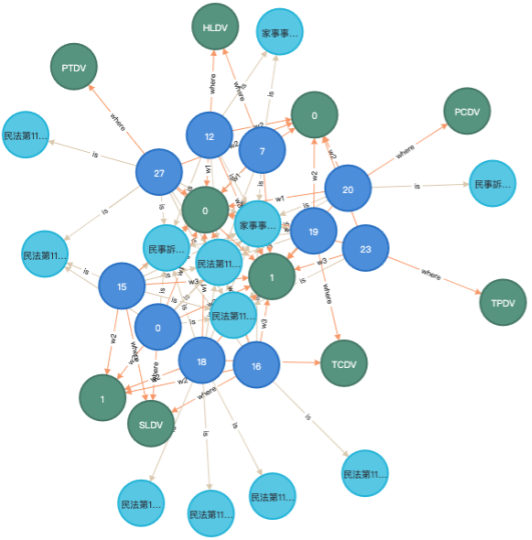} 
  \caption{Judgement Relationship Network Graph with LDA topics} 
\end{figure}
\vspace{-0.5cm} 

And the total work flow can be shown like this figure 5. 

\begin{figure}[H] 
  \centering 
  \includegraphics[width=\linewidth]{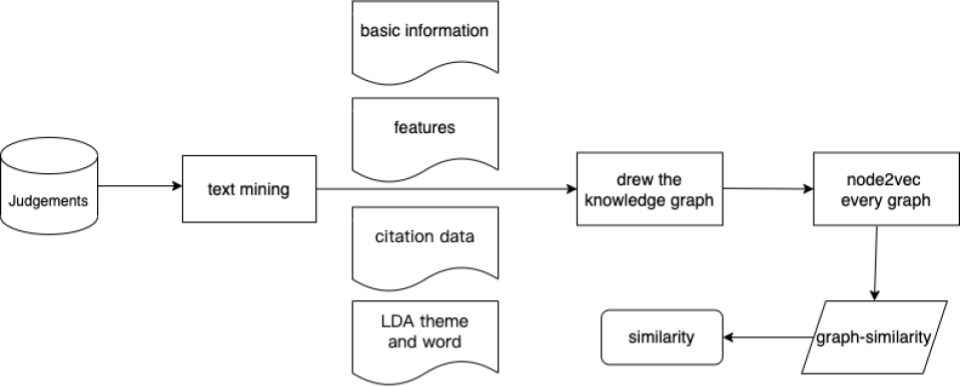} 
  \caption{Research Flow} 
\end{figure}
\vspace{-0.5cm} 

\section{Judgment2Vec} 

In practice, we tried to make this process easier by graph analytics; then, we drew the knowledge graph by the “case-article” relationship. The primary thought is that the case used more law articles in this judgment, more similar. Then, we could make a similar rank for each case and even build a graph-similarity matrix. The result can be compared to two methods: labeled-similarity and graph-similarity and analyzed by human experts.

With this graph, we could check and have more clues to calculate the “judgment features.” Furthermore, the process can be automatically realized, and we could enter more data samples to build a more excellent legal knowledge graph/ network.

We use the Neo4j to build a node(judgment) vector\cite{r3}. Each judgments can be presented as a vector, we name it as Judgment2Vec. Then we could use this vector to calculate cosine similarity of Judgment2Vec.

\begin{figure}[H] 
  \centering 
  \includegraphics[width=\linewidth]{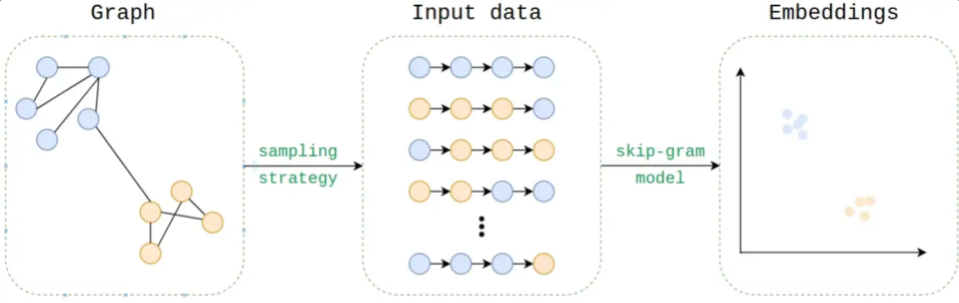} 
  \caption{ Graph Embedding[3]} 
\end{figure}
\vspace{-0.7cm} 

\section{Research Result and Discussion} 

In the end, we can get two “differences in similarity.” Because of pairwise comparison, there are 13,571 “similar referee combinations.” The minimum can be the same, and the maximum can differ to 0.83. Please see Figure 7 below.

\begin{figure}[H] 
  \centering 
  \includegraphics[width=\linewidth]{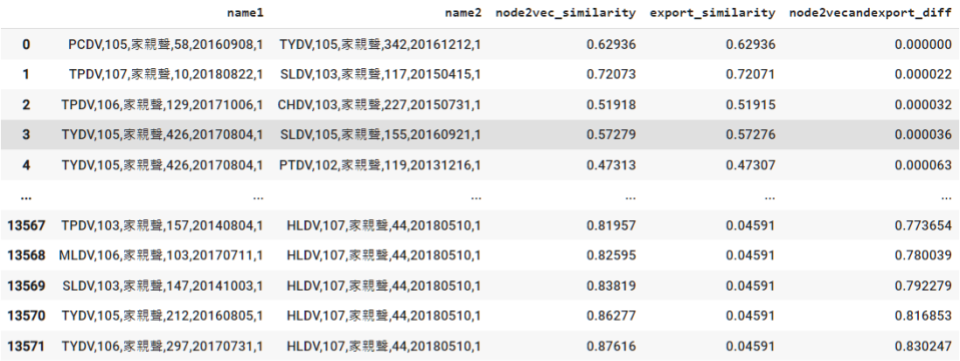} 
  \caption{Similarity Pair of Two Methods} 
\end{figure}
\vspace{-0.7cm} 

However, the median difference is 0.19, while the mean is 0.21, see table 1. Because the literature\cite{r4} only revealed the degree of correlation between algorithms but did not reveal the relevant information with expert evaluation. Therefore, this study currently establishes the state-of-art of automatic calculation and human expert cognition.

\begin{table}[H]
\caption{Describe Statistic of Similarities}
\resizebox{\linewidth}{15mm}{
\begin{tabular}{|l|l|l|l|}
\hline
\begin{tabular}[c]{@{}l@{}}index\\ n=13572\end{tabular} & node2vec\_similarity & export\_similarity & node2vecandexport\_diff \\ \hline
mean                                                    & 0.6193               & 0.5300             & 0.2154                  \\ \hline
std                                                     & 0.1007               & 0.2221             & 0.1464                  \\ \hline
min                                                     & 0.3341               & 0.0459             & 0.0000                  \\ \hline
25\%                                                    & 0.5475               & 0.3500             & 0.0971                  \\ \hline
50\%                                                    & 0.6200               & 0.5192             & 0.1951                  \\ \hline
75\%                                                    & 0.6929               & 0.7151             & 0.3122                  \\ \hline
max                                                     & 0.9231               & 1.0000             & 0.8302                  \\ \hline
\end{tabular}
}
\end{table}
\vspace{-0.5cm} 

The contributions of this study are: 1. Established a legal judgment marking table of the expert system, which can be used as related research. 2. Established an automated process that imitates human marking so that using graph embedding, it can imitate humans to find similar judgments. 3. Established concrete realization results in real-world data; the matrix diagram established in Figure 8 can be used as a retrieval system. Most legal searches will look for “most similar,” so in practice, the main objects of attention should be those parts with minor differences (for example, judgments with differences below o.1), and less attention will be paid to those parts with significant differences.

In this stage, we could expand this method to another judgments corpus dataset, which does not have a human label yet. Shortly speaking, input a text dataset, and output a knowledge graph and similarity matrix of those judgments. It could be extended as a recommendation or search system. The contributions of this study are: 1. Established a legal judgment marking table of the expert system, which can be used as related research. 2. Established an automated process that imitates human marking so that using graph embedding, it can imitate humans to find similar judgments. 3. Established concrete realization results in real-world data; the matrix diagram established in Figure 8 can be used as a retrieval system. Most legal searches will look for “most similar,” so in practice, the main objects of attention should be those parts with minor differences (for example, judgments with differences below 10\%), and less attention will be paid to those parts with significant differences. 

In terms of research limitations, this paper finds that preprocessing techniques may have a significant impact; for example, the standardization of some compensation amounts, or the “addition/removal” of some legal variables, may affect the degree of similarity. This issue requires more in-depth cooperation with legal experts, and we hope that the automated method established in this study can assist legal practitioners in finding similar judgments.








\bibliography{ref} 
\end{document}